\documentclass[anon]{ceurart}
\usepackage{listings}
\usepackage{listings}
\begin{document}

\copyrightyear{2026}
\copyrightclause{Copyright for this paper by its authors.
  Use permitted under Creative Commons License Attribution 4.0
  International (CC BY 4.0).}

%% Conference information
\conference{AI4SciSci 2026: 3rd International Workshop on
Artificial Intelligence for the Science of Science,
co-located with ACM/IEEE-CS JCDL 2026,
October 13, 2026, Frisco, Texas, USA}

%% Title
\title{FutureLens: Longitudinal Topic Patterns in Scholarly Text and Future-Work References}

% \tnotetext[1]{You can use this document as the template for preparing your
%   publication. We recommend using the latest version of the ceurart style.}

%% Anonymous Author Block for Double-Blind Review
% \author{Anonymous Author(s)}
%% Authors
\author[1]{Ibrahim Al Azher}
\ead{iazher1@niu.edu}

\author[1]{Hamed Alhoori}
\ead{alhoori@niu.edu}

%% Affiliation
\address[1]{Department of Computer Science,
Northern Illinois University,
DeKalb, IL, USA}

\begin{abstract}
Future-work statements offer a record of the questions researchers choose to leave open. Using them to describe a field, however, requires separating thematic change from changes in document coverage, topic assignment, and scholarly writing. FutureLens examines these measurement issues in a recovered snapshot of 5,655 ACL papers published during 2012--2022, including 4,005 with supplied future-work reference text. We compare non-negative matrix factorization (NMF) and latent Dirichlet allocation (LDA), aggregate topic presence at the paper level, and examine normalized prevalence, exploratory trend statistics, assignment stability, and agreement between input-derived text and references. Across available seeds, NMF has higher mean normalized pointwise mutual information (0.1387 versus 0.1280) and higher pairwise assignment stability, while LDA has greater keyword diversity (0.8900 versus 0.7750). Seed-zero results show increasing prevalence of broad model-training and question-related vocabulary, alongside substantial citation and document-structure content. Mean input/reference topic-set overlap is 0.5191 for NMF and 0.4924 for LDA; temporal rank agreement varies widely and does not imply agreement in prevalence levels. These findings support a descriptive account of the available corpus, with explicit qualifications for missing references, incomplete runs, and unverified reference provenance. The contribution is an empirical measurement study of longitudinal scholarly signals, not a new topic-modeling algorithm or a forecasting system. 
Code and data are available at
\url{https://github.com/IbrahimAlAzhar/FutureLens_longitudinal} and \url{https://huggingface.co/datasets/IbrahimAlAzhar/FutureScope_longitudinal}.
\end{abstract}

% https://anonymous.4open.science/r/FutureLens_longitudinal-86F4/README.md  

\maketitle

\section{Introduction}

Scientific papers document both completed work and unfinished questions.
Their future-work statements occupy a distinctive rhetorical position:
authors identify unresolved problems, possible extensions, or resources
they would like to develop. Previous studies have analyzed these statements
to characterize proposed research directions and differences across NLP
tasks~\cite{qian2021,zhang2023}. Aggregating these statements can help
readers inspect how a research community describes its open agenda.
This complements analyses of publication text and citation
patterns~\cite{anderson2012}, but it does not establish which proposed
directions are subsequently pursued.

The distinction matters for Science of Science. A rise in mentions can
reflect more publications, better recovery of future-work text, a change
in terminology, or a change in the underlying research agenda. Topic
interpretation also requires care: statistical model performance does
not necessarily correspond to human judgments of
interpretability~\cite{chang2009}. Automatic coherence measures provide
useful evidence about topic quality~\cite{lau2014}, but a cluster with
a plausible title may still contain routine experimental language rather
than a coherent scientific problem. Even when two text sources exhibit
similar temporal rankings, they may differ substantially in how often
they mention the corresponding topic. A useful longitudinal analysis
must expose these alternatives before interpreting its output as
evidence about scientific priorities.

% FutureLens addresses this measurement problem through a descriptive
% study of ACL papers from 2012--2022. We compare input-derived scholarly
% text with supplied future-work references in a common topic space within
% each model and seed. We analyze topics using non-negative matrix
% factorization (NMF)~\cite{lee1999} and latent Dirichlet allocation
% (LDA)~\cite{blei2003}. We report paper-normalized prevalence, trend
% statistics, seed stability, and source agreement as separate quantities.
% The study uses a recovered experimental snapshot: available outputs
% support NMF and LDA analyses, while several planned components have no
% usable results. This boundary defines the paper's empirical scope.

FutureLens investigates how reliably future-work statements
can characterize the evolution of research themes within a
scientific community. We examine 5,655 ACL papers published
between 2012 and 2022, including 4,005 papers with available
future-work references. Using non-negative matrix factorization
(NMF) ~\cite{lee1999} and latent Dirichlet allocation (LDA) ~\cite{blei2003}, we analyze the
reliability of topic representations, identify longitudinal
patterns in expressed future directions, and examine their
relationship with the themes of current scholarly research.
Rather than treating changes in topic prevalence as direct
evidence of shifting scientific priorities, we investigate
how corpus coverage, topic-model variability, and textual
artifacts constrain their interpretation. Our analysis is
descriptive and does not establish whether proposed research
directions are subsequently pursued.

Our study addresses three research questions.
\textbf{RQ1: How reliably can topic models identify and characterize
research themes in scholarly publications and future-work statements?}
We examine differences between NMF and LDA in topic coherence,
keyword diversity, and assignment stability across random seeds.
\textbf{RQ2: How have research themes expressed in ACL future-work
statements evolved from 2012 to 2022, and how do corpus coverage
and topic content affect their interpretation?}
We analyze normalized topic prevalence and longitudinal trends
while accounting for incomplete reference coverage and potential
textual artifacts.
\textbf{RQ3: To what extent do future-work statements reflect
the themes of current scholarly research?}
We investigate topic overlap at the paper level and temporal
agreement between scholarly input text and future-work references,
distinguishing similarities in trend direction from differences
in topic prevalence.

Our contributions are threefold. First, we provide a comparative
evaluation of NMF and LDA for characterizing scholarly research
themes, distinguishing topic coherence, keyword diversity, and
assignment stability from semantic validity. Second, we present
a coverage-aware longitudinal analysis of future-work statements
in ACL publications from 2012 to 2022, identifying prominent
thematic patterns while examining the effects of reference
availability, topic content, and temporal variation. Third, we
analyze the relationship between current scholarly research
and expressed future directions through paper-level topic overlap
and temporal source agreement. Together, these contributions
provide an empirical framework for interpreting longitudinal
scholarly signals while recognizing the limitations of incomplete
coverage, model sensitivity, and unverified reference provenance.

\section{Related work and positioning}
\textbf{Longitudinal topic analysis of scientific literature.}
Computational histories of NLP already use publication text to study changes in research activity. Anderson et al.~\cite{anderson2012} analyze the ACL's intellectual history, while Pramanick et al.~\cite{pramanick2023} study paradigm shifts through a framework involving causal discovery and inference. Our narrower focus is the measurement relationship between input-derived scholarly text and supplied future-work references; our recovered results do not support causal inference.

\textbf{Future-work statements.}
Qian et al.~\cite{qian2021} use future-work sentences to explore
research trends across NLP tasks. Zhang et al.~\cite{zhang2023}
develop recognition and classification methods, examine
future-work content, and compare it with abstracts of
subsequent research. Yang et al.~\cite{yang2025} identify
emerging topics through entity novelty in future-work
sentences. FutureGen~\cite{al2025futuregen} employs an
LLM-based retrieval-augmented generation (RAG) approach
to generate future-work suggestions from scholarly texts.
Future-work analysis and its connection to research
evolution are therefore established research directions.
FutureLens examines a complementary measurement question:
how model sensitivity, reference coverage, and source
disagreement affect longitudinal interpretation.
We distinguish same-year trajectory agreement from
subsequent uptake and make no claim to identify emerging
entities or predict their adoption.

\textbf{Topic models and interpretation.}
LDA models documents through latent topic mixtures~\cite{blei2003}; NMF decomposes nonnegative representations into additive components~\cite{lee1999}. Dynamic topic models explicitly allow topic representations to change with time~\cite{blei2006}. Here, a fixed representation within each descriptive fit makes annual prevalence comparable, at the cost of imposing a retrospective vocabulary. Neither this design nor the use of topic labels constitutes a new modeling algorithm.

LLM-assisted interpretation also has direct precedents. LimTopic combines topic modeling and language-model-based interpretation for limitations in scientific articles~\cite{limtopic}. We treat generated titles as reading aids attached to recovered keyword lists. They do not change topic assignments and are not independent evidence that a topic is meaningful. This separation is consistent with the broader distinction between model fit and human interpretability~\cite{chang2009}. Automatic coherence measures can assist evaluation~\cite{lau2014}, but our study does not substitute them for an annotation experiment.

% Our contribution is a corpus-specific measurement study: we jointly examine coverage, assignment consistency, and source agreement to establish which descriptive conclusions the available evidence supports. This complements prior work on future-work extraction and emerging-topic identification.
Prior studies examine future-work identification,
emerging topics, and longitudinal scholarly trends.
FutureLens complements these approaches by jointly
examining topic-model reliability, corpus coverage,
and agreement between current research and supplied
future-work references. This measurement-oriented
analysis clarifies how these factors constrain
the interpretation of longitudinal scholarly signals. 

% Existing studies have investigated future-work identification, emerging research topics, and longitudinal changes in scientific literature. However, interpreting future-work statements as
% indicators of research evolution requires careful consideration of topic-model reliability, corpus coverage, and the relationship between current research and proposed future directions. FutureLens examines these interconnected measurement challenges within a common empirical setting. Unlike approaches focused primarily on extracting future-work statements or identifying emerging topics, our study evaluates the reliability of topic representations, characterizes coverage-aware temporal patterns, and analyzes agreement between scholarly input text and future-work references. This provides a complementary, measurement-oriented perspective on the interpretation of longitudinal scholarly signals.

\section{Corpus and evidence boundaries}
\subsection{Observed coverage}
% The recovered ACL scope contains 5,655 analyzed papers across eleven publication years, with 4,005 supplied references and 1,650 missing references. Overall reference availability is 70.82\%. Table~\ref{tab:coverage} reports the annual denominators; these are counts in the analyzed snapshot, not total ACL publication counts. The 2015 sample contains only 33 papers, so its prevalence estimates are particularly sensitive to individual records. The snapshot does not establish why annual coverage is uneven.
We focus on ACL publications to examine longitudinal research
patterns within a relatively coherent scientific community,
where changes in research tasks, methodologies, and terminology
can be studied within a shared disciplinary context.
The 2012--2022 observation period reflects the available
recovered experimental snapshot rather than the complete
historical or contemporary coverage of ACL publications.

\begin{table}[tb]
\centering
\caption{Analyzed ACL coverage. Reference availability is computed from the recovered counts; missing references are retained in the all-paper prevalence denominator.}
\label{tab:coverage}
\begin{tabular}{rrrrr}
\toprule
Year & Papers & References & Missing & Available (\%)\\
\midrule
2012 & 187 & 106 & 81 & 56.68\\
2013 & 328 & 240 & 88 & 73.17\\
2014 & 286 & 205 & 81 & 71.68\\
2015 & 33 & 19 & 14 & 57.58\\
2016 & 328 & 227 & 101 & 69.21\\
2017 & 302 & 220 & 82 & 72.85\\
2018 & 381 & 260 & 121 & 68.24\\
2019 & 753 & 550 & 203 & 73.04\\
2020 & 861 & 624 & 237 & 72.47\\
2021 & 1,165 & 814 & 351 & 69.87\\
2022 & 1,031 & 740 & 291 & 71.77\\
\midrule
Total & 5,655 & 4,005 & 1,650 & 70.82\\
\bottomrule
\vspace{-14pt}
\end{tabular}
\end{table}

\subsection{Text sources and recovered runs}
We distinguish \emph{input-derived activity text} from \emph{supplied reference text}. The former is a textual proxy for the research discussed in a paper, not an external observation of scientific activity. The inspected pipeline constructs ACL input from the abstract and six supplied text fields. Its activity transformation removes exact reference sentences and selected future-work cues. References come from the supplied \texttt{Future\_Work} field; the report labels their source \texttt{acl\_supplied} and records no fallback. This processing does not establish whether references were manually extracted, automatically extracted, or edited upstream. We therefore avoid the term ``ground truth.'' The report contains three run variants associated with Gemma, Mistral, and Qwen topic labeling. All lack completion markers. NMF has five available seeds in each; LDA has four in the Gemma and Mistral variants and three in the Qwen variant. Matching numerical outputs are not three independent experiments: these language models only title topics in this pipeline. We use the first listed, Gemma-labeled snapshot as the reporting reference, without pooling variants. Its five NMF and four LDA seeds determine the method summaries. Detailed descriptive tables use seed zero, the lowest seed with such outputs, rather than the highest-scoring seed.
No usable BERTopic outputs, forecasting records, silhouette values, or human judgments are recovered. Missing values remain unavailable. We do not import numerical results from the earlier, differently configured 2012--2024 manuscript into this 2012--2022 comparison. Appendix~\ref{sec:provenance} explains the evidence and implementation boundary.

% \section{Preliminary BERTopic exploration.}
% An initial FutureLens analysis applied BERTopic independently to yearly collections of future-work text from ACL papers spanning 2012–2024, with an LLM generating descriptive titles for the resulting keyword clusters. The yearly summaries highlighted themes such as machine translation, multilingual processing, and model evaluation, illustrating the potential of future-work text for exploring a community’s expressed research directions. However, independently fitted yearly clusters do not establish consistent topic identities across time, and their raw frequencies depend on annual corpus size. These exploratory observations motivate the present analysis using a common topic space within each fit and paper-normalized prevalence. Because the corpus period and experimental configuration differ, the earlier BERTopic outputs are kept separate from the recovered NMF/LDA comparison. 

\section{Preliminary BERTopic Exploration}
An earlier FutureLens analysis applied BERTopic
independently to yearly ACL future-work collections
from 2012--2024, identifying themes such as machine
translation, multilingual processing, and evaluation.
However, independently fitted yearly clusters lack
consistent topic identities, and raw frequencies
are sensitive to corpus size. These limitations
motivate our present analysis using a shared topic
space within each fit and normalized prevalence.
The preliminary results are not combined with
the current NMF/LDA comparison.

\section{Problem formulation and analysis design}
\subsection{Common topic space and paper-level presence}
Let $D_t$ denote analyzed papers in year $t$, with $N_t=|D_t|$. For paper $i$, let $A_i$ denote activity text, $R_i$ supplied reference text, and $m_i$ indicate whether $R_i$ is available. Each model and seed defines its own topic set $\mathcal{K}=\{0,\ldots,19\}$. The inspected implementation fits the descriptive topic space on deduplicated sentence units pooled from activity and reference text across years. The recovered stability tables compare 1,168,951 identical training units. These units are not additional papers or independent temporal observations.

% For NMF, the implementation uses TF--IDF features, random initialization, and up to 400 iterations. LDA uses count features, batch learning, and up to 20 iterations. Both use English stop-word removal, a maximum vocabulary of 20,000 features, and $K=20$ topics. This compares conventional representations and algorithms jointly; no topic-count sensitivity study is available. 
For NMF, the implementation uses TF--IDF features, random initialization, and up to 400 iterations.
LDA uses count features, batch learning, and up to 20 iterations. Both models use English stop-word
removal and a maximum vocabulary of 20,000 features.

We fix $K=20$ for both models to ensure comparable
thematic granularity. This is a controlled
experimental choice, not an empirically optimized
topic count. As different topic counts may alter
thematic structures and longitudinal patterns,
our findings remain conditional on this specification.
The comparison also varies feature representations
and modeling algorithms.

Each sentence receives its maximum-weight topic; zero-feature units receive label $-1$. A paper's topic set $T_i^S$, for source $S\in\{A,R\}$, is the union of nonnegative sentence labels. Define $z_{ikt}^{S}=\mathbb{1}[k\in T_i^S]$. A topic is counted at most once per paper, although a paper can contain many topics. Consequently, topic prevalences need not sum to one. Longer text offers more opportunities to acquire topics even under this binary aggregation.

The global descriptive fit keeps topic identifiers fixed across years within a model and seed. It does not align identifiers across NMF and LDA or across seeds. Because later text contributes to the fit, a topic containing a later term such as \emph{BERT} may also absorb earlier, more generic model-training sentences. Its 2012 prevalence cannot be interpreted as evidence that the later technology was present then.

% \begin{figure}[tb]
% \centering
% \figplaceholder{19mm}{Figure placeholder: input text and supplied references $\rightarrow$ sentence units $\rightarrow$ a common NMF or LDA topic space $\rightarrow$ paper-level topic sets $\rightarrow$ prevalence, stability, and source agreement. Place LLM titles on a separate interpretive branch.}
% \caption{Descriptive analysis design. Each fit spans the observed years; source comparisons use that fit's topic vocabulary. Titles are annotations, not an additional measurement stage.}
% \label{fig:pipeline}
% \end{figure}

\subsection{Two denominators answer different questions}
The recovered trend tables use all analyzed papers:
\begin{equation}
 P_{kt}^{A}=\frac{\sum_{i\in D_t}z_{ikt}^{A}}{N_t},\qquad
 P_{kt}^{R,\mathrm{all}}=\frac{\sum_{i\in D_t:m_i=1}z_{ikt}^{R}}{N_t}.
\label{eq:all}
\end{equation}
The second quantity measures \emph{observed reference-topic coverage among all analyzed papers}. Missing references contribute no observed topic occurrence, but are not verified topic absences. Let $M_t=\sum_{i\in D_t}m_i$, $c_t=M_t/N_t$, and $P_{kt}^{R,\mathrm{obs}}$ be prevalence among papers with references. Then
\begin{equation}
 P_{kt}^{R,\mathrm{all}}=c_t P_{kt}^{R,\mathrm{obs}}.
\label{eq:coverage}
\end{equation}
This identity separates reference availability from topic prevalence conditional on availability. Normalization removes the direct effect of corpus size; it does not remove missingness bias. We retain the reported all-paper series when presenting its saved statistics and use available-reference endpoint calculations only as explicitly identified denominator checks.

For temporal source agreement, the pipeline instead restricts both sources to the same papers, $D_t^*=\{i\in D_t:m_i=1\}$. We denote these paired prevalences by $\widetilde P_{kt}^{A}$ and $\widetilde P_{kt}^{R}$, each with denominator $M_t$. These are different estimands from the all-paper activity series and reference coverage in Equation~\ref{eq:all}.

\subsection{Quality, stability, and trends}
Recovered coherence uses binary word co-occurrence over saved training text units. For two topic keywords $u,v$, normalized pointwise mutual information is
\begin{equation}
 \operatorname{NPMI}(u,v)=\frac{\log[p(u,v)/(p(u)p(v))]}{-\log p(u,v)}.
\end{equation}
The report reconstructs NPMI from saved text and diversity from saved keywords. Diversity is the fraction of distinct words across the top-ten keyword lists for all 20 topics. Both are intrinsic summaries, not measures of factual accuracy or historical validity. Across seeds, we report means and sample standard deviations using only available observations.

Assignment stability is adjusted Rand index (ARI)~\cite{hubert1985} between seed pairs on identical training units, including the outlier label. ARI compares partitions without requiring matching topic identifiers. The pairs share seeds and are dependent; we summarize their range without treating them as independent replications. Assignment stability also does not establish stability of a particular named topic's trajectory.

The Sen slope summarizes pairwise annual prevalence changes~\cite{sen1968}:
\begin{equation}
 \widehat\beta_k=\operatorname{median}_{a<b}
 \frac{P_{k,t_b}-P_{k,t_a}}{t_b-t_a}.
\end{equation}
The implementation uses Kendall rank association with year and Benjamini--Hochberg (BH) adjustment within each method/source trend table~\cite{bh1995}. We report saved slopes in percentage points (pp) per year and saved adjusted values $q$. The eleven annual observations, uneven sample sizes, temporal dependence, and dependence between topics limit inference. A small $q$ is an exploratory association under this procedure, not evidence of a causal transition. The reported intervals are not propagated uncertainty from parsing, model fitting, or missingness; we do not relabel them as bootstrap validation.

\subsection{Source overlap and temporal agreement}
For each paper with a reference, topic-set overlap is
\begin{equation}
 J_i=\frac{|T_i^A\cap T_i^R|}{|T_i^A\cup T_i^R|}.
\end{equation}
An empty union is undefined. All 4,005 available-reference rows are scored. Because the sources describe different aspects of a paper and reference provenance is unverified, Jaccard is neither extraction precision nor recall.

For each topic, temporal agreement comprises Spearman correlation $\rho_k$ between the eleven paired annual prevalence values and the mean absolute prevalence gap,
\begin{equation}
 G_k=\frac{1}{11}\sum_{t=2012}^{2022}|\widetilde P_{kt}^{A}-\widetilde P_{kt}^{R}|.
\end{equation}
Correlation describes similarity in temporal ordering; $G_k$ describes separation in levels. Both are contemporaneous and descriptive. No lagged target or held-out future observation enters this comparison.

\section{Results}
\subsection{RQ1: A trade-off between coherence and diversity}
Table~\ref{tab:quality} shows higher recovered mean NPMI for NMF and higher keyword diversity for LDA. NMF's pairwise ARI range, 0.4183--0.7429, lies above LDA's 0.2161--0.2633 in this snapshot. Both have a recovered outlier rate of 0.0163. Under the inspected assignment rule, this is the share of zero-feature training units; it should not be read as a comparison of density-based outlier detection.

\begin{table}[tb]
\centering
\caption{Recovered method comparison in the reporting snapshot. Quality values are mean $\pm$ sample SD across available seeds. ARI ranges cover ten NMF and six LDA seed pairs on 1,168,951 identical units, including label $-1$. No significance test between methods is available.}
\label{tab:quality}
\begin{tabular}{lcc}
\toprule
Quantity & NMF & LDA\\
\midrule
Available seeds & 5 (0--4) & 4 (0--3)\\
Topics per seed & 20 & 20\\
NPMI & $0.1387\pm0.0033$ & $0.1280\pm0.0016$\\
Keyword diversity & $0.7750\pm0.0061$ & $0.8900\pm0.0227$\\
Outlier rate & $0.0163\pm0.0000$ & $0.0163\pm0.0000$\\
Pairwise ARI range & 0.4183--0.7429 & 0.2161--0.2633\\
\bottomrule
\end{tabular}
\vspace{-7pt}
\end{table}

Neither model is uniformly superior. NMF exhibits higher
coherence and assignment stability, whereas LDA achieves
greater keyword diversity. However, these intrinsic
metrics do not establish semantic validity. In particular,
the low NPMI variation across seeds coexists with
substantial assignment differences, especially for LDA.
Moreover, assignment stability does not guarantee
consistent longitudinal trajectories for individual
topics. Thus, our findings characterize model-level
reliability, while topic-specific temporal robustness
remains unverified.

\subsection{RQ2: Longitudinal patterns and their interpretation}
Table~\ref{tab:trends} summarizes selected longitudinal patterns from
seed zero, including increasing and declining themes,
non-significant changes, and citation-related artifacts.
These patterns describe shifts in topic prevalence within
the analyzed corpus rather than verified changes in
scientific priorities. Topic descriptions are based on
recovered keywords, and selection is illustrative rather
than exhaustive. Reported q values retain their
original within-table multiple-testing adjustment.

\begin{table}[tb]
\centering\small
\caption{Selected reference-topic patterns, 2012--2022. Percentages use all analyzed papers: $N_{2012}=187$, $N_{2022}=1,031$. Changes and slopes are in pp and pp/year. Short descriptions paraphrase recovered keywords; they are not independently validated labels.}
\label{tab:trends}
\begin{tabular}{llrrrrr}
\toprule
Topic & Keyword-based description & 2012 & 2022 & Change & Sen & BH $q$\\
\midrule
NMF 5 & Models / training & 20.86 & 43.16 & 22.31 & 2.2379 & .0005\\
NMF 8 & Datasets / questions & 20.32 & 40.25 & 19.93 & 1.2543 & .0017\\
NMF 7 & Citations / years & 28.88 & 47.43 & 18.55 & 1.2993 & .0009\\
NMF 0 & Language / translation & 37.43 & 51.70 & 14.26 & 0.6634 & .2532\\
NMF 15 & Attention / representation & 46.52 & 56.94 & 10.41 & $-0.0327$ & 1.0000\\
LDA 19 & Model training / fine-tuning & 21.93 & 43.55 & 21.62 & 2.6370 & .0003\\
LDA 14 & Questions / generation & 10.70 & 30.94 & 20.25 & 1.9543 & .0001\\
LDA 4 & Syntax / parsing & 31.02 & 28.61 & $-2.40$ & $-1.2310$ & .0473\\
LDA 9 & Features / POS / older years & 34.22 & 16.59 & $-17.64$ & $-2.7994$ & .0038\\
\bottomrule
\end{tabular} 
\vspace{-9pt}

\end{table}

\textbf{Model-training and question-related vocabulary.}
NMF topic 5 rises from 20.86\% to 43.16\% in references, with slope 2.2379 pp/year (saved 95\% interval: 1.5331--2.7825) and $q=.0005$. Its activity prevalence rises from 67.91\% to 93.02\%, with slope 2.7806 and $q=.0003$. LDA topic 19 shows a broadly related pattern, rising from 21.93\% to 43.55\% in references and from 67.38\% to 93.89\% in activity text. The keyword overlap supports a qualitative comparison; it does not establish a one-to-one topic alignment between methods. Moreover, Question-related components also increase: reference prevalence rises by 19.93 pp for NMF topic 8 and 20.25 pp for LDA topic 14. These components combine dataset, question, generation, or document vocabulary. They support a change in the prevalence of these broad lexical themes, but not the more specific assertion that authors increasingly propose a particular type of question-answering dataset.

\textbf{Coverage changes the size of the contrast.}
The all-paper NMF topic 5 counts are 39 references in 2012 and 445 in 2022. Dividing by the available-reference counts gives $39/106=36.79\%$ and $445/740=60.14\%$, an endpoint increase of 23.34 pp. This is a direct calculation from the reported counts, not a new fitted trend. Conversely, LDA topic 4 changes from $58/106=54.72\%$ to $295/740=39.86\%$ among available references, compared with an all-paper decline of only 2.40 pp. Reference availability rose between these endpoints and can mask a conditional decline. Neither calculation resolves selection bias among papers with missing references, and the saved $q$ values cannot be transferred to these conditional series.

\textbf{Endpoint change is not monotonic change.}
NMF topic 15 has a 10.41 pp endpoint increase but an approximately zero, slightly negative Sen slope and $q=1.0000$. NMF topic 0 also rises between endpoints, yet its trend has $q=.2532$. These examples show why a last-minus-first difference should not be described as sustained growth. For LDA topic 4, a small endpoint decline coexists with a more pronounced negative Sen slope. Intermediate years affect the slope, although their trajectories cannot be reconstructed from endpoints alone.

\textbf{Some prominent components are textual artifacts.}
NMF topic 7 contains \emph{et}, \emph{al}, and publication years; LDA topic 1 has a similar vocabulary. Their reference increases are evidence of changing citation-related content, not new research agendas. LDA topic 9 mixes features and POS with older year tokens, so its decline cannot be attributed entirely to a decline of a technical paradigm. Other components contain figure, table, section, or web-address vocabulary. We retain such components in the reported statistical families but identify them as a validity concern. A fluent generated title cannot turn these mixtures into a clean scientific category.

% \begin{figure}[tb]
% \centering
% \figplaceholder{18mm}{Figure placeholder: annual paired-source prevalence for NMF topics 5, 8, 13, and 14, with annual reference availability in a separate panel. Populate from the saved annual trajectory tables; do not interpolate the recovered endpoints.}
% \caption{Planned visualization of levels and temporal agreement. The recovered report supplies summary statistics but not the full annual trajectories needed to render these panels faithfully.}
% \label{fig:trajectories}
% \end{figure}

\subsection{RQ3: Source agreement depends on the measure}
Mean paper-level Jaccard overlap is 0.5191 for NMF and 0.4924 for LDA over 4,005 scored papers. These are moderate descriptive overlaps in the respective learned topic spaces. Their difference is not an extraction-accuracy gain, and no uncertainty estimate or paired significance test is available. Topic granularity and source length can affect Jaccard even if the underlying text is unchanged.

\begin{table}[tb]
\centering
\caption{Selected temporal source-agreement summaries for seed zero. Both sources are restricted to papers with available references in each of eleven years. Gap values are the reported fractions multiplied by 100, hence pp. Examples illustrate agreement and disagreement; no correlation significance tests are claimed.}
\label{tab:agreement}
\begin{tabular}{llrr}
\toprule
Topic & Keyword-based description & Spearman $\rho$ & Mean gap (pp)\\
\midrule
NMF 5 & Models / training & .8000 & 33.48\\
NMF 8 & Datasets / questions & .7636 & 43.61\\
NMF 13 & Words / embeddings & .9364 & 42.42\\
NMF 14 & Work / related / future & .2364 & 5.20\\
NMF 3 & Data / training / corpus & $-.3727$ & 35.19\\
LDA 4 & Syntax / parsing & .9182 & 35.84\\
LDA 14 & Questions / generation & .8636 & 30.80\\
LDA 19 & Model training / fine-tuning & .8818 & 38.23\\
LDA 8 & Data / annotation & $-.3235$ & 32.75\\
\bottomrule
\end{tabular}
\end{table}

Table~\ref{tab:agreement} demonstrates that temporal rank correlation and level agreement answer different questions. NMF topic 13 has $\rho=.9364$ but a mean absolute gap of 42.42 pp. NMF topic 14 has a much smaller gap, 5.20 pp, but $\rho=.2364$. Close prevalence levels need not move together; similarly ordered series can remain far apart. These gaps use paired papers and cannot be reconstructed by subtracting the all-paper endpoint percentages in Table~\ref{tab:trends}. Temporal agreement is also heterogeneous. Model-training components have positive correlations in both methods, whereas NMF topic 3 and LDA topic 8 have negative correlations. These latter components both include broad data-related vocabulary but are not assumed to match. Their behavior cautions against a single claim that future-work references mirror activity text. Shared fitting, overlapping vocabulary, and residual source dependence also limit the independence of this comparison, even after deterministic removal of exact reference sentences.

\section{Discussion and Limitations}

FutureLens reveals increasing prevalence of broad
model-training and question-related themes, alongside
declining syntax-related content and substantial
textual artifacts. These findings highlight three
distinct measurement considerations: intrinsic
topic quality, assignment stability, and agreement
between current research and future-work statements.
None independently establishes the validity of
longitudinal research trends.

Several limitations constrain interpretation.
First, the analyzed ACL corpus covers 2012--2022
with uneven annual representation, including only
33 papers in 2015. Moreover, 29.18\% of papers
lack future-work references, and missingness may
be systematic. Normalized prevalence does not
eliminate this potential selection bias or establish
generalizability beyond the analyzed corpus.

Second, reference provenance remains unverified,
and missing references may reflect either absent
future-work statements or incomplete extraction.
Differences in text length, cue-based filtering,
and citation-related artifacts may affect topic
assignments. Without human validation, topic
interpretations remain provisional.

Third, the fixed $K=20$ specification and
retrospective fitting constrain temporal
interpretation. Although aggregate stability
is assessed across seeds, detailed trends rely
on seed zero and lack topic-count sensitivity
validation. The eleven annual observations also
limit statistical inference.

Finally, the recovered experiments provide no
usable forecasting or comparable BERTopic results.
Our source comparisons capture contemporaneous
thematic relatedness rather than subsequent
research uptake. Future work should assess
reference validity, missingness, and model
robustness, and compare future-work themes
at year $t$ with published research at $t+1$,
using existing research prevalence as a baseline.

\section{Conclusion}
FutureLens examines the reliability and longitudinal
interpretation of research themes expressed in 5,655
ACL papers and their available future-work references.
NMF and LDA exhibit distinct coherence, diversity, and
stability characteristics, while normalized prevalence
reveals broad thematic changes alongside citation and
document-structure artifacts. Paper-level overlap and
temporal agreement further demonstrate that future-work
statements and current research are related but not
interchangeable sources of evidence. These findings
highlight the importance of corpus coverage, topic-model
reliability, and source-specific interpretation when
examining scholarly evolution. FutureLens provides a
descriptive foundation for future studies investigating
whether expressed research directions are subsequently
reflected in published scientific activity.

\section*{Declaration on Generative AI}

During the preparation of this work, the authors used ChatGPT/Claude for minor sentence polishing, grammar and spelling checks, and rephrasing to improve clarity and readability. All AI-assisted text was reviewed and edited by the authors, who take full responsibility for the publication's content.

\bibliography{sample-ceur}

\appendix
\section{Evidence Provenance and Reproducibility Boundary}
\label{sec:provenance}

All experimental values in the main text originate in the supplied
recovered-results report, generated on 11 September 2026. The report
explicitly identifies partial snapshots, reconstructs NPMI, diversity,
outlier rate, and assignment stability where possible, and preserves
saved trend and agreement tables. It prefers readable CSV over JSON
and does not count the same table twice. This paper adds only transparent
arithmetic conversions, such as annual reference availability and
selected conditional endpoint percentages.

The corresponding no-extraction pipeline archive was inspected to
clarify feature construction, assignment, trend tests, and the paired
denominator used for trajectory agreement. This source inspection
documents the implementation; it is neither a rerun nor proof that the
exact inspected revision produced every saved file. The report does
not provide a checksum linking the executed source code to the inspected
archive. Confirming that linkage, software versions, and upstream
reference provenance remains necessary for end-to-end reproducibility.
Statements about implementation describe the inspected archive;
numerical findings describe the recovered snapshot. No missing
experiment is inferred from configuration settings.

The preliminary analysis supplies the motivation for examining
prospective scholarly statements. Its independently fitted yearly
BERTopic models, 2012--2024 corpus, extraction and refinement metrics,
embedding comparisons, hierarchical categories, and trend figure are
not combined with the recovered 2012--2022 results. Combining these
outputs would conflate different datasets and measurement procedures.
Any discussion of the preliminary analysis therefore provides
exploratory context rather than evidence for the NMF/LDA comparison.

\begin{table}[htbp]
    \centering
    \caption{Availability of recovered runs. Variants are reported
    separately. The rows do not represent additional independent
    papers or independent evaluations of labeling models.}
    \label{tab:run-availability}
    \begin{tabular}{lccc}
        \toprule
        Labeling variant & NMF seeds & LDA seeds & Completion marker \\
        \midrule
        Gemma   & 0--4 & 0--3 & Absent \\
        Mistral & 0--4 & 0--3 & Absent \\
        Qwen    & 0--4 & 0--2 & Absent \\
        \bottomrule
    \end{tabular}
\end{table}

% Remove the following paragraph once the figures are finalized.
The two figure boxes are explicit production placeholders. The first
can be replaced with a diagram of the documented descriptive workflow.
The second requires the saved annual paired trajectory tables;
endpoint and correlation summaries cannot determine the underlying
curves. No graph has been fabricated from those summaries. The
placeholders contain no experimental evidence beyond the surrounding
text and tables.

\end{document}